\documentclass[12pt]{article}

\usepackage{hyperref}
\usepackage{natbib}
\usepackage{amsthm} %For theorems
\usepackage{amsfonts} %For mathbb
\usepackage{amscd} % Diagrams
\usepackage{amsmath} %Align environment
\usepackage{graphicx}
\usepackage{epstopdf}
\usepackage{pgf}
\usepackage{array, multirow}
\usepackage{caption}
\usepackage{subcaption}

\usepackage{pgfpages}
\usepackage{authblk}

\usepackage{amsmath}
\usepackage{natbib}
\newcommand{\inlinecode}[1]{\texttt{#1}}
\newcommand{\m}[1]{\mathcal{#1}}
\newcommand{\up}[1]{\text{#1}}
\newcommand{\mat}[1]{\text{\bf #1}}

\renewcommand{\d}{\ \mathrm{d} }

\newcommand{\dd}[1]{\frac{\mathrm{d}}{\mathrm{d} #1}}

\title{How to solve the stochastic partial differential equation that gives a Mat\' ern random field using the finite element method}
\author{Haakon Bakka \texttt{bakka@r-inla.org}}
\begin{document}
\maketitle

\begin{abstract}
This tutorial teaches parts of the finite element method (FEM), and solves a stochastic partial differential equation (SPDE).
The contents herein are considered ``known'' in the numerics literature, but for statisticians it is very difficult to find a resource for learning these ideas in a timely manner (without doing a year's worth of courses in numerics).

The goal of this tutorial is to be pedagogical and explain the computations/theory to a statistician. % in a concise way.
% (or other outside the field of numerics).
If you find you do not understand one of the parts, let me know, and I will update it.
This is not a practical tutorial; there is little computer code, and no data analysis.
This work is based on the paper by \citet{lindgren2011spdeapproach}.
\end{abstract}

\section{Introduction}

In this tutorial, we detail how to implement functions that build precision matrices (inverse covariance matrices) from SPDEs (stochastic partial differential equations), using the FEM (finite element method).

In the main part of this tutorial we solve the differential equation
% I use we for me and you, and people in general, but I for ''only me''.
\begin{align}
k u - \nabla^2 u = f(x,y) 
\label{eq-spde-diff}
\end{align}
where $f(x,y)$ is deterministic, $\nabla^ 2 = \left(\frac{\partial^ 2}{\partial x^ 2}, \frac{\partial^ 2}{\partial y^ 2} \right)$, and $k$ is a constant.
After we have solved this deterministic equation, only a small step is needed to make it stochastic (by making $f$ stochastic).
Solving this equation means specifying a discretization into a finite vector
$$\vec u = (u_i)_{i=1}^N. $$ % = u(x_i, y_i), $$
This discretization is ``using a basis'' in FEM terminology.
And then, we want to find matrices
$$\mat J, \quad \mat D $$
such that the differential equation becomes
\begin{align}
\label{eq-matrices}
k \mat J \vec u + \mat D \vec u = \vec f.
\end{align}
Notice how the minus sign becomes a plus sign.\footnote{This 
	is intuitively because
the laplacian $\nabla^2$ represents outflow, and we want to subtract the out-flow, so the negative laplacian is the positive term.}
We use arrows to denote vector representations (i.e.\ discretisations) of functions, which are very high dimensional vectors, but we will use boldface for other vectors.

Using the FEM means that $u_i$ not only represents $u(x_i, y_i)$, which is common for ``finite differences'' and pointwise approximations, but also the function's values on a finite element around $(x_i, y_i)$.
A finite element is also known as a hat-function; its precise definition depends on the mesh, and is found in section \ref{sec-element-hat-fun}.

Before we continue, I suggest the reader becomes familiar with the $\nabla$ operator if this is not already the case.
The $\nabla$ applied to a function is the gradient, see \url{https://en.wikipedia.org/wiki/Gradient}, while the $\nabla$ operator applied to a vector (for example the gradient vector) is the divergence of that vector, see \url{https://en.wikipedia.org/wiki/Divergence}.
An example of how to compute the gradient and divergence can be found at
\url{https://youtu.be/qN5wxhHHCIM}.

%This tutorial contains illustrative examples, and example code. 
This is not a paper, and some of the text is not written to be mathematically rigid, 
but to explain as clearly as possible what is going on. 
This includes some repetition, and it includes some ``fluffy text'' 
that is supposed to paint an intuitive image.
Only a minimal level of mathematical details are included.
However, for the subject of this tutorial, that turns out to be quite a lot of concepts and equations.

\section{The weak formulation and the space where the equation is defined}
For technical reasons, equation \eqref{eq-spde-diff} needs to be written using integrals.
The weak formulation (integral formulation) of equation \eqref{eq-spde-diff} is  
\begin{equation}
\label{eq-weak}
k\int u \cdot v \d \Omega - \int\nabla ^2 u\cdot v\d \Omega = -\int f\cdot v \d \Omega \qquad \text{for all } v.
\end{equation}
where $u, v, f$ are functions of $x$ and $y$.\footnote{
Technically, they have to be in the sobolev space $\up H^1(\Omega)$, which is a precise definition of being ''sufficiently nice''.}
The integrals are over the domain $\Omega$, if nothing else is specified.
The $\Omega$ is the entire spatial area where equation \eqref{eq-spde-diff} is to be solved.
The$\d \Omega$ is just a way of writing$\d x \d y$, the two-dimensional integral, without having to write two integration signs.
The dot $\cdot$ is ordinary multiplication, the sign is used to clearly separate the different terms. 
When the dot is used between two vectors, that is vector multiplication (inner product).

By applying Green's first identity\footnote{
\url{https://en.wikipedia.org/wiki/Green's\_identities}
}, we obtain
\begin{equation}
\label{eq-greens}
k\int u \cdot v \d \Omega +\int\nabla u\cdot \nabla v\d \Omega - \int_{\partial \Omega}v \cdot (\nabla u\cdot {\bf \hat n})\d \sigma = \int f\cdot v \d \Omega \quad \text{for all } v.
\end{equation}
Note that $\nabla u = \nabla u(x,y)$ is a function producing 2-dimensional vectors.

To be clear, if a function $u = u(x,y)$ solves this equation for any function $v=v(x,y)$, then, per definition, $u$ is a solution to equation \eqref{eq-spde-diff}.
The problem with equation \eqref{eq-spde-diff} is that it is often not mathematically meaningful, and so, when we write that equation, we actually want to solve equation \eqref{eq-weak}.
Note also that we do not need the second derivative of $u$ to exist for equation \eqref{eq-greens} to make sense.

\subsection{Neumann boundary conditions}
The Neumann boundary condition says that the derivative is zero on the boundary. 
Which derivative? 
There are several directions to move in!
The derivative is taken to be zero when you move away from the boundary.
To write this in mathematics, we note that $\dd x$ is the derivative in direction $[1, 0]$, and $\dd y$ is the derivative in direction $[0, 1]$.
So we can define the derivative in a vector direction ${\bf n}$ as a combination of these.
Writing down the Neumann boundary condition then is
\begin{align}
\label{eq-neumann}
\frac{\partial u(x,y)}{\partial {\bf n}}|_{\partial \Omega_N} &= 0,
\end{align}
where the vertical bar gives notation for ``at the boundary''.

In calculus (of several dimensions), it is well known that this expression is equivalent to
$$ \nabla u\cdot {\bf \hat n} = 0, $$
where ${\bf \hat n}$ is the normal vector on the boundary (${\bf \hat n}$ depends on where you are on the boundary). This is very useful for simplifying our equation \eqref{eq-greens}.

A intuitive explanation of this boundary condition is that the outflow is zero, when considering the derivative to be the flow.
This makes sense since the derivative is a vector.
Another name for this boundary condition is the ``natural boundary condition''. 
This highlights the fact that it eliminates a term from our equation \eqref{eq-greens} and makes our life simpler.

\subsection{Continuing with the equation}
We now have to solve
\begin{equation}
\label{eq-nice}
k \int u \cdot v \d \Omega +\int\nabla u\cdot \nabla v\d \Omega
  = \int f\cdot v \d \Omega,
\end{equation}
which we rewrite into
\begin{align}
\label{eq-JDl}
kJ(u,v) + D(u,v) &= l(v), \forall v \in X,
\end{align}
where $J(u,v) = \int u v$ etc. 
Later we will deal with one part at a time.
The naming of these functions of functions $J$ and $D$ is the same as the matrices in equation \eqref{eq-matrices} on purpose.
Functions of functions that produce a number (such as $J$ and $D$) we call functionals, and functions of functions that produce new functions we call transformations, but this naming has no impact on the mathematical definitions (one could call everything ``functions'').
This far in the tutorial we have only rewritten the equations into a suitable form;
we have yet to do any approximations.

\section{What are hat-functions (elements)?}
\label{sec-element-hat-fun}
The hat-functions are the elements in the FEM.
Also known as the basis functions (since they give a basis for the space of solutions). 
We call them hat-functions since they look like sharp hats.

\begin{figure}
\centering
\includegraphics[width=.6\linewidth]{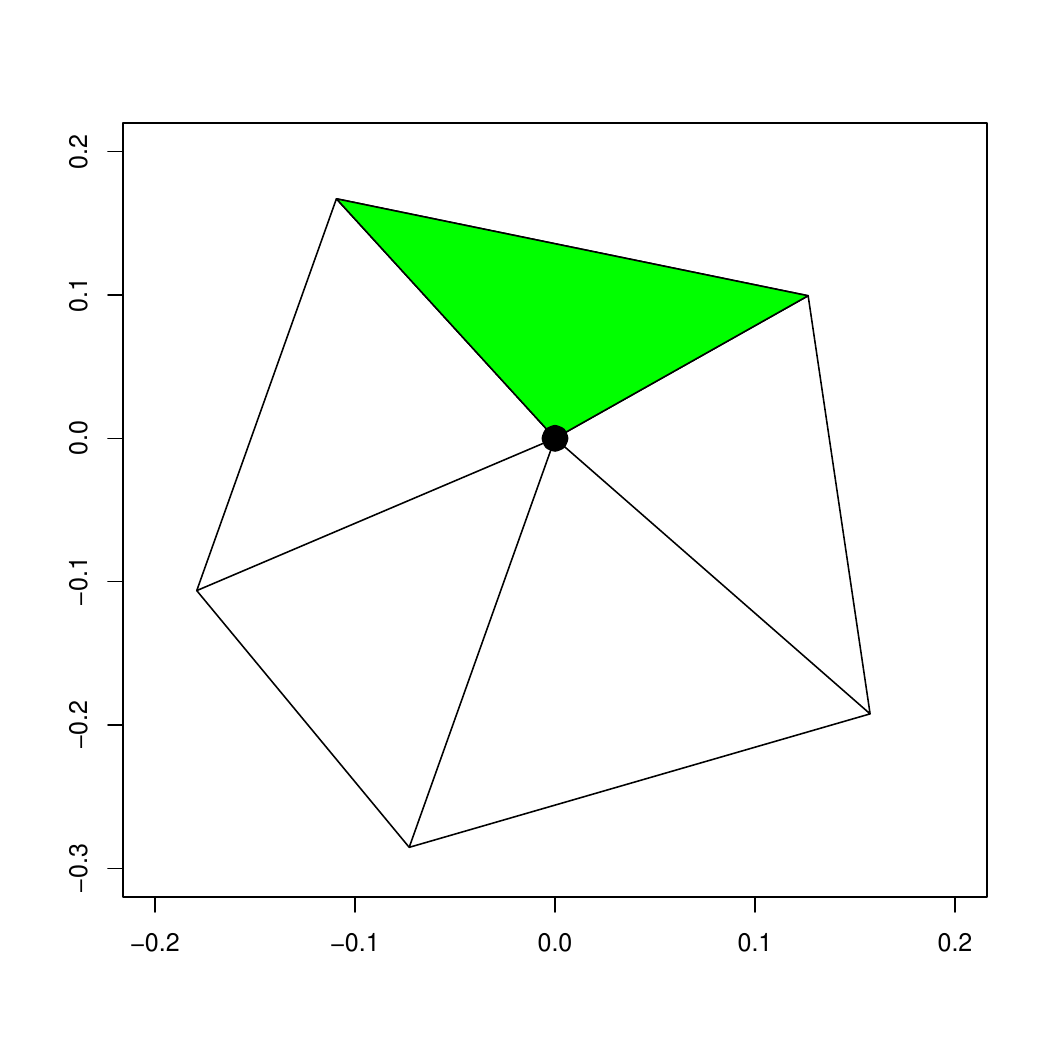}
\caption{One node and its neighbouring triangles. 
One triangle is highlighted in green.}
\label{fig-center-node-and-triangles}
\end{figure}

To define a single hat-function $\phi_i$, we must first pick a node in the mesh, and investigate all the triangles around the node.
Let us move the center node to $(0,0)$ for convenience.
See Figure  \ref{fig-center-node-and-triangles}.
Note that there can be any number of triangles around the point.
The hat-function is 0 outside of the neighbouring triangles.\footnote{
	In this tutorial we only consider linear FEM. In general, the basis functions can be much more complex.}

We define the hat-function separately on each triangle, as a linear function that is 1 at the central node (0,0), and decreases linearly to zero at each of the two other corners (nodes) in the triangle.
Knowing the values of a linear function at three points uniquely determines the function.
We will not write down the expression here.
(An example hat-function can be seen in the bottom right of Figure \ref{fig-linapprox}.)

When we have defined the hat-function on all these triangles, we note that the function has the same values on each edge, and so is consistent.
We also note that the function is zero at all the edges that are not connected to the central node.
Hat-functions at neighbouring nodes will overlap.

\section{The space of (approximate) solutions}
We will now define the space of finite element solutions $X_h$.
Specifying this space of approximate solution functions, is equivalent to specifying a mesh in our case.\footnote{
In this entire tutorial, I assume that you want to use \emph{linear} finite element method.
This is not the only possible choice. 
You can use linear elements for ``anything'', as the mesh becomes finer, it will converge to the true solution.
Non-linear FEM may converge in a different way, quicker or slower, and are more complicated to implement.
}
Let us explain this in detail.

As an aside, for the reader who knows about Hilbert spaces, we note that $X$ is an infinite dimensional Hilbert space, and $X_h$ is a finite dimensional (Hilbert) subspace.
And so, the act of approximating can be considered as a projection from the big space into the finite dimensional space.
It is this mathematical framework that ensures finite element approximations are very nice in practice.

\begin{figure}
\centering
\includegraphics[width=.45\linewidth]{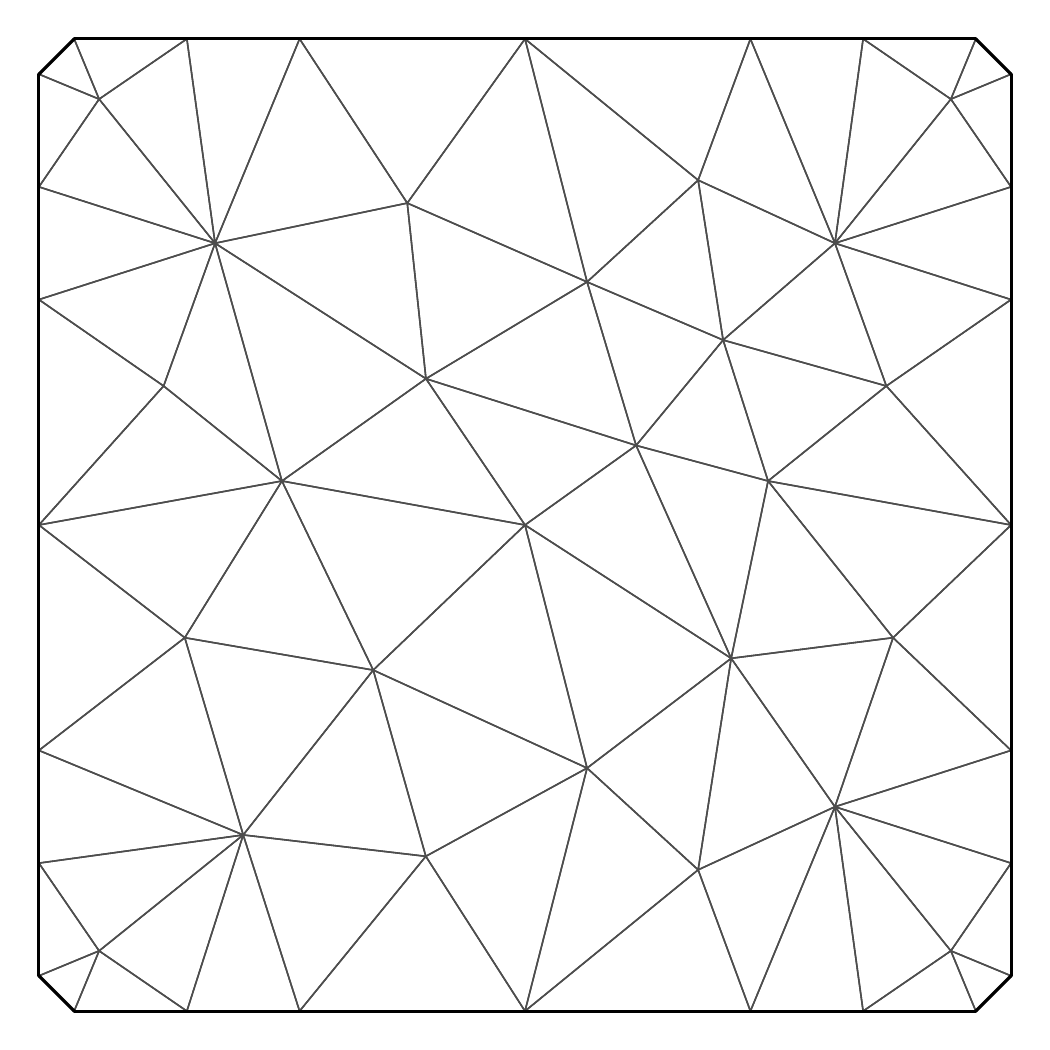}
\caption{Example mesh. 
We note how irregular these triangles are. 
This irregularity is one of the big advantages with the FEM method, as your domain can have any kind of shape, and you may do more detailed approximations in some places than in others.}
\label{fig-mesh1}
\end{figure}

A mesh is a triangularisation of the domain, which means to take the spatial study area (the domain) and divide it up into small triangles.
You can set up a simple mesh by using the \texttt{R}-code:
\begin{verbatim}
> library("INLA")
> mesh = inla.mesh.2d(loc = matrix(c(0,0,1,1,0,1,0,1), 4, 2), 
    max.edge = 0.3)
> plot(mesh)
\end{verbatim}
% pdf("mesh1.pdf"); par(mai=c(0,0,0,0)); plot(mesh, main=""); dev.off()
See Figure \ref{fig-mesh1} for this mesh.

\begin{figure}
\centering
\includegraphics[width= .45\textwidth]{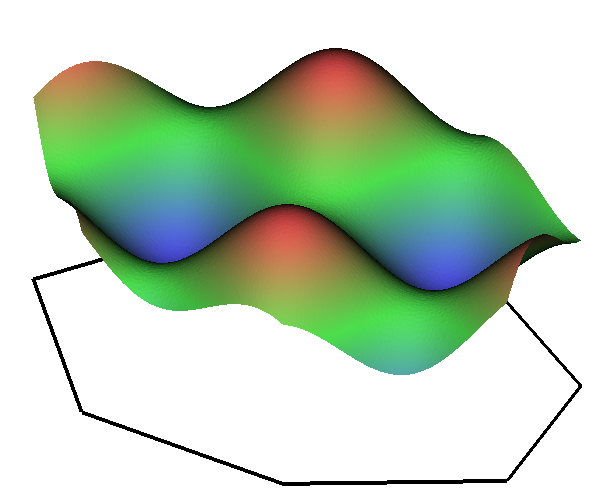}
\includegraphics[width= .45\textwidth]{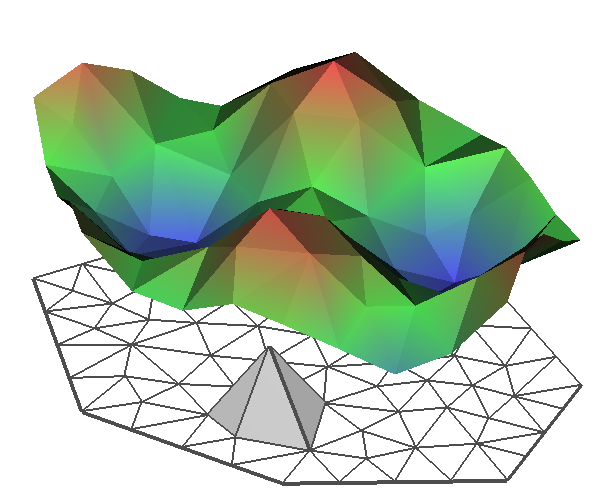}
\caption{Linear approximations. 
The function to the left is not approximated.
The function to the right is piecewise linear.
Underneath it you see the mesh, and you see an example hat-function (element) growing out of the mesh.
The mesh defines ``piecewise linear'', and the hat-function is the simplest piecewise linear function.
What you need to realise is that the surface on the right is a linear combination of these hat-functions. 
Also, we note that the function and its approximation looks very similar everywhere, and not just in some points, an improvement over many other approximation methods.
}
\label{fig-linapprox}
\end{figure}

The (approximate) solutions are piecewise linear functions with respect to this mesh.
This means that on each triangle they can be written as a linear function ($a+bx+cy$).
See figure \ref{fig-linapprox} for an example.
The meaning of piecewise linear depends on the choice of pieces (i.e.\ triangles), so from now on, whenever we say piecewise linear, we mean ``with respect to the mesh''.

Any piecewise linear function can be written as a linear combination of hat-functions. 
%TODO
If you have trouble with seeing this connection, between piecewise linear functions and a linear combination of hat-functions, that is quite common. 
We include appendix \ref{app-1d} where we look at this connection in 1 dimension. 
After having understood it in 1 dimension, you can come back to this section.

\section{Approximating a known function}
\label{sec-approx-known-fun}
For approximating a function by a piecewise linear function, it is essential that we do \emph{not} use
$$ f_i = f(x_i, y_i)$$
as this is the \emph{wrong} way to approximate a function with a piecewise linear function.
What we need to do is ask ``how many hat-functions to you consist of''.
This is done by finding a piecewise linear approximation $m(s)$ satisfying
$$\int m(x) \phi_i(x) = \int f(x) \phi_i(x) \text{for all } i. $$
Write $m$ as a linear sum of basis functions, and we can expand the left hand side as
$$\sum_j m_j  \int \phi_j(x) \phi_i(x), \text{  in vector form:  } \mat J \vec m $$
by using the $\mat J$ matrix developed later in this tutorial.
The length of the vector $\vec f$ is the same as the total number of hat-functions, which is the same as the number of nodes in the mesh.

As a special case, for any piecewise linear function $f(x_i, y_i)$ 
we have that $f_i = f(x_i, y_i)$ at the nodes, 
because it is already an approximation, 
and thus, finding the value $f(x_i, y_i)$ happens to be equivalent to 
asking ``how many hat-functions do you consist of''.

\section{Reference triangle and the mapping $\up T$}
\label{sec-ref-tri-and-T}
The arbitrary triangles scattered in space are not fun to work with.
Here we define a transformation to a reference triangle, with corners in origo (0,0) and in (1,0) and (0,1).
All computations will be done on this reference triangle, keeping in mind the transformation.

Pick one specific triangle, with corners
$$\bf p_1, p_2, p_3, $$
represented as columns of x- and y-values. 
Let $\vec z$ be the coordinates in the space where we define the mesh and the functions, and let $\vec \eta$ be the coordinates in the reference triangle (a new space). 
In this section we use the arrow-notation to represent an arbitrary vector in 2 dimensions, while boldface represents fixed vectors.
We use the transformation
\begin{align}
\vec z &= \text T {\vec \eta} + {\bf p_1},\\
\text T &= \left[\begin{array}{cc}
{\bf (p_2-p_1)} & {\bf (p_3-p_1)}
\end{array}\right], \\
\left[\begin{array}{c}
z_1 \\ 
z_2
\end{array}\right] &= 
\left[\begin{array}{cc}
{\bf (p_2-p_1)} & {\bf (p_3-p_1)}
\end{array}\right]
\left[\begin{array}{c}
\eta_1 \\ 
\eta_2
\end{array}\right] 
+ {\bf p_1}.
\end{align}
Clearly $E$ sends $(0,0)$ to $\bf p_1$, $(1,0)$ to $\bf p_2$. and $(0,1)$ to $\bf p_3$. 
The triangle with corner nodes $(0,0), (0,1)$ and $(1,0)$ is our reference domain $\tilde \Omega$. 
The vector ${\bf z}$ is the coordinates in domain $\Omega$ while ${\vec \eta} = [\eta_1, \eta_2]^\top$ is the reference coordinates of $\tilde \Omega$. 

To work with the hat-function $\phi_i$, we choose one of the triangles, say $t$, where it is defined, transform $t$ to the reference triangle by letting $\bf p_1$ to be the node $i$, and on the reference triangle, this part of $\phi_i$ is 
$$1-x-y. $$

If we want to integrate a function $g$ over any triangle, we use
\begin{align}
\int_{\Omega} g({\vec z}) \d {\vec z} &= |\text{detT}|\int_{\tilde{\Omega}} g(\text T{\vec \eta} + {\bf p_1})\d {\vec \eta}
\end{align}
and integrate over the reference triangle instead.

This $\phi_i$ is not the only hat-function that is nonzero in this triangle, 
there is one which is centered around the top node (0, 1), with function value $y$ on this triangle, 
and another which is centered around the right node (1,0), with function value $x$ on this triangle.

\subsection{Approximating the solution} \label{sec:approxsol}
We want to approximate the solution to equation \eqref{eq-JDl}, with piecewise linear functions. 
We want to find $u \in X_h$ with
\begin{align}
\label{eq-jdl2}
kJ(u,v) + D(u,v) &= l(v), \forall v \in X_h.
\end{align}

We now write $u$ and $v$ as
\begin{align*}
u &= \sum_i u_i \phi_i \\
v &= \sum_j v_j \phi_j
\end{align*}

The integrals $J$ and $D$ and $l$ are all linear in each of their arguments.
This implies that solving equation \eqref{eq-jdl2} for any $v$ that is a linear combination of $\phi_j$'s, is equivalent to solving it for any single hat-function $\phi_j$.
Showing this is straight forward, and therefore not included here.

Further, because of linearity,
\begin{align}
\label{eq-jdl3}
kJ(u,\phi_j) + D(u,\phi_j) &= l(\phi_j), \forall j \\
kJ\left(\sum_i u_i \phi_i,\phi_j \right) + D \left(\sum_i u_i \phi_i,\phi_j\right) &= l(\phi_j), \forall j \\
\sum_i  \left( kJ \left(\phi_i,\phi_j \right) + D \left( \phi_i,\phi_j\right) \right) u_j &= l(\phi_j), \forall j
\end{align}
and we rewrite this as matrix equations
\begin{align}
\label{eq-matrix2} k \mat J \vec u &+ \mat D \vec u = \vec f, 
\end{align}
where
\begin{align}
\mat J_{i,j} &= J(\phi_i, \phi_j) 
= \int \phi_i \cdot \phi_j  \d \Omega \\
\mat D_{i,j} &= D(\phi_i, \phi_j) 
= \int \nabla \phi_i \cdot \nabla \phi_j  \d \Omega \\
f_j &= l(\phi_j) = \int f\cdot \phi_j \d \Omega.
\end{align}

Ending up with the matrix equation \eqref{eq-matrix2} is the goal of the FEM discretisation.
The only part that remains is to compute the elements in the different matrices.
The $f_j$ here is not the same as the linear approximation of $f$ in Section \ref{sec-approx-known-fun}, but only the first step to compute that linear approximation.

\section{Computing $\mat D_{i,j} = D(\phi_i,\phi_j)$}
With $\phi_i(z(\eta))= \tilde \phi_i(\eta )$, note that
\begin{align}
\d ^2 z = |\det(J_{\vec \eta}(\vec z))| \d^2\eta,
\end{align}
where $J_{\vec \eta}(\vec z) = \text T $ is the derivative of $\vec z = E(\vec \eta )= \text T\vec \eta + \vec p_1$ with respect to $\eta$. 
Here, $\d ^2 z$ represents $\d z_1 \d z_2$.
Note that T depends on triangle index $t$.
When we write $\nabla$ this is implicitly $\nabla_\eta$, i.e.\ the derivative in the normal space where the mesh and equations are defined.

\begin{align}
D(\phi_i,\phi_j) &= \int_{\Omega} \nabla\phi_i \cdot \nabla\phi_j \d^2 { z}\\
 &= \sum_{t \in V_{i,j}} \int_{\Omega_t} \nabla_z \phi_i \cdot \nabla_z \phi_j \d^2 { z}\\
 &= \sum_{t \in V_{i,j}}\ |\text{detT}|\int_{\tilde{\Omega}_t} \nabla_z\phi_i \cdot \nabla_z\phi_j \d^2 { \eta} \\
  &= \sum_{t \in V_{i,j}}\ |\text{detT}| \int_{\tilde{\Omega}_t} (\nabla_\eta \phi_i \text T^{-1}) \cdot (\nabla_\eta\phi_j \text T^{-1}) \d^2 { \eta} \\
  &= \sum_{t \in V_{i,j}}\ |\text{detT}| \int_{\tilde{\Omega}_t} (\nabla_\eta \tilde \phi_i \text T^{-1}) (\nabla_\eta\tilde\phi_j \text T^{-1}) ^\top \d^2 { \eta} \\
  &= \sum_{t \in V_{i,j}}\ |\text{detT}| \int_{\tilde{\Omega}_t}  \nabla_\eta \tilde \phi_i (\text T ^\top \text T)^{-1} (\nabla_\eta\tilde\phi_j)^\top \d^2 { \eta} 
  \end{align}
where $V_{i,j}$ is the set of indexes of all triangles having ${\bf p}_i$ and ${\bf p}_j$ as one of their corner nodes,  
$\Omega _t$ is triangle number $t$, 
and $\tilde{\Omega}_t$ is the reference triangle.

In the case where $i=j$ the $V_{i,j}$ contains a few triangles, but if $i \neq j$ there are only 1 or 2 triangles in $V_{i,j}$.

The derivative $\nabla_\eta \tilde \phi_i$ is constant over the triangle, and there are only 3 possible values for $\nabla_\eta \tilde \phi_i$, namely
\begin{align*}
(-1,-1) & \text{ if } \phi_i \text{ is situated at } \vec p_1 \\
(1,0) & \text{ if } \phi_i \text{ is situated at } \vec p_2 \\
(0,1) & \text{ if } \phi_i \text{ is situated at } \vec p_3,
\end{align*}
see the last paragraph before Section \ref{sec:approxsol}.
This makes the integrand constant, and since the area of the reference triangle $\tilde{\Omega}_t$ is $\frac{1}{2}$ we can write
\begin{align}
D(\phi_i,\phi_j) 
  &= \sum_{t \in V_{i,j}}\ |\text{detT}| \frac{1}{2}  (\nabla_\eta \tilde \phi_i) (\text T ^\top \text T)^{-1} (\nabla_\eta\tilde\phi_j)^\top
\end{align}
We used this expression in the function \inlinecode{dt.fem.laplace} in \inlinecode{INLA:::inla.barrier.fem} in the \texttt{R}-library \texttt{INLA}.
%For the Barrier model, see \url{https://haakonbakka.bitbucket.io/btopic107.html}.
%For the functions mentioned throughout this tutorial, see
%\url{https://haakonbakka.bitbucket.io/functions-barriers-dt-models-march2017.R}.

The matrix $\mat D$ is usually called the stiffness matrix in FEM literature.

\section{Computing $\mat J_{i,j} = J(\phi_i,\phi_j)$}
Computing this matrix is less complicated, fortunately.

\begin{align}
J(\phi_i,\phi_j) &= \int_{\Omega} \phi_i \cdot \phi_j \d^2 { z}\\
&= \int_{\Omega} \phi_i \cdot \phi_j \d^2 { z}\\
&= \sum_{t \in V_{i,j}}\ |\text{detT}| \int_{\tilde{\Omega}_t} 
 \tilde \phi_i \cdot \tilde \phi_j \d^2 { \eta},
\end{align}
where the integrals are easy to compute, becoming
\begin{align*}
\int_{\tilde{\Omega}_t} 
 \tilde \phi_i \cdot \tilde \phi_j \d^2 { \eta}
 &= 1/12 \text{ for } i=j \\
 &= 1/24 \text{ for } i \neq j
\end{align*}

This is used in the function \inlinecode{dt.fem.identity} in \inlinecode{INLA:::inla.barrier.fem}.

\section{Solving the differential equation sparsely}
There is one important detail we have not mentioned about solving the differential equation, a detail which is often not stated in numeric literature, because it is the only reasonable approach.
That detail is sparse matrices.

A matrix is usually represented in a computer as lists of all the elements.
FEM matrices are usually very big, possibly with millions of rows. 
Fortunately, most of the elements of these matrices are zero.
Because of this, the matrices are stored in a sparse way.
Essentially, they are represented as a set of indices $(i,j)$ and values $v$.
Where there is a zero element in the matrix, there is usually no index $(i,j)$ pointing to that element.
In ths way, zeroes are neither stored, nor used when computing sums or products or solving linear systems.

The matrix equation
$$ (k \mat J + \mat D) \vec u = \vec f $$
is solved through sparse matrix algebra, using a sparse matrix library.
This completes the goal of solving the differential equation.

\section{Solving the stochastic equation}
The only part that remains now is to deal with a stochastic right hand side $f$.
We will assume it is Gaussian, so that any finite-dimensional approximation $\vec f$ can be represented by a covariance matrix.
Conveniently, all the computations for the left hand side is the same.
We need only define a covariance matrix $\mat C$ for $\vec f$, and the solution becomes
\begin{align}
(k \mat J + \mat D) \vec u = \vec f  \\
Cov\left((k \mat J + \mat D) \vec u \right)= C \\
(k \mat J + \mat D) Cov\left( \vec u \right) (k \mat J + \mat D) = C
\end{align}
since the matrix $(k \mat J + \mat D)$ is symmetric.
We often write $\mathcal W$ instead of $f$ to highlight that it is considered some kind of white noise.

For computational reasons we cannot use dense matrices, and so, we use the precision matrix (inverse covariance matrix) instead.
We denote by $Q_u$ the precision matrix for vector $u$,
\begin{align}
Q_u = (k \mat J + \mat D) Q_f (k \mat J + \mat D).
\label{eq-QaQa}
\end{align}
The only part of this tutorial that remains is to choose a precision matrix $Q_f$ for the driving noise, and to ensure that it is sparse.

We use 
\begin{align}
(Q_{f,i,i})^{-1} &= Cov(f_i, f_i)
  =  \int_{\Omega} \phi_i \d \Omega \\
&= \sum_{t \in V_{i,j}}\ |\text{detT}| \int_{\tilde{\Omega}_t} 
 \tilde \phi_i  \d^2 { \eta}, \\
 &= \sum_{t \in V_{i,j}}\ |\text{detT}| \cdot c,
\end{align}
where all other entried of $Q_f$ are zero.
We can compute the value $c$, but $c$ is a multiplicative constant for the entire system of equations, so you can always rescale at the end.
We use this equation in \inlinecode{dt.fem.white} in \inlinecode{INLA:::inla.barrier.fem}.

The reason for this choice of covariance is that the continuous Gaussian white noise has variance equal to the area, and the integral of the hat-function corresponds to this area.

Instead of this iid Gaussian noise, we can use the precision matrix from any kind of process, 
including the precision matrix from solving another SPDE.
For a discussion on some of the spatial models based on solving SPDEs, see \cite{bakka2018spatial}.

\section{Connection to the Mat\' ern field}
In the title, we refer to the SPDE 
\begin{align}
k u - \nabla^2 u = \m W
\end{align}
where $\m W$ is continuously indexed white noise, to represent the Mat\' ern field, 
in the sense that the solution (with some restrictions) 
is the Mat\'ern random field (with a certain range and smoothness).
The exact connection between the SPDE and the Mat\' ern field is found in \citet{lindgren2011spdeapproach} (first two sections), including references to the original sources.

For completeness, we mention the two rules of thumb that in practice gives you a good representation of the Mat\' ern field when using equation \eqref{eq-QaQa}.
With ``range'' we refer to the empirical range in \citet{lindgren2011spdeapproach}.
The Neumann boundary condition we imposed when formulating the weak solution gives an inaccurate representation of the Mat\' ern field near the boundary, hence we recommend having a distance between the boundary and any inference/prediction location of 1/2-1 range.
The FEM approximation depends on the longest edge length of any triangle in the interior mesh (the mesh where we do predictions and inference);
this ``max.edge'' should be 1/10-1/5 of the range.
However, in some applications we break these two rules, usually when a low computational cost is more important than representing the Mat\' ern field well.
When we break these rules, we always get a proper and well defined spatial model, the only downside is that it is not the Mat\' ern field.

\section{Code examples for spatial modeling}
We provide a simple simulation-inference code at
\url{https://haakonbakkagit.github.io/btopic122.html}
and a simple code example at
\url{https://haakonbakkagit.github.io/btopic108.html}.
An introduction to coding spatial models in R-INLA can be found at \url{http://www.r-inla.org/spde-book}, together with a large collection of advanced examples.

\section{Acknowledgements}
Thanks to David Bolin and H\aa{}vard Rue for feedback before the first version of the tutorial.
Special thanks to Parker Trostle for detailed feedback on the code and the second version.

\bibliography{local}

% New appendix:
\appendix

\section{Mesh and piecewise linear in 1D}
\label{app-1d}
\begin{figure}
	\centering
	\includegraphics[width=.45\textwidth]{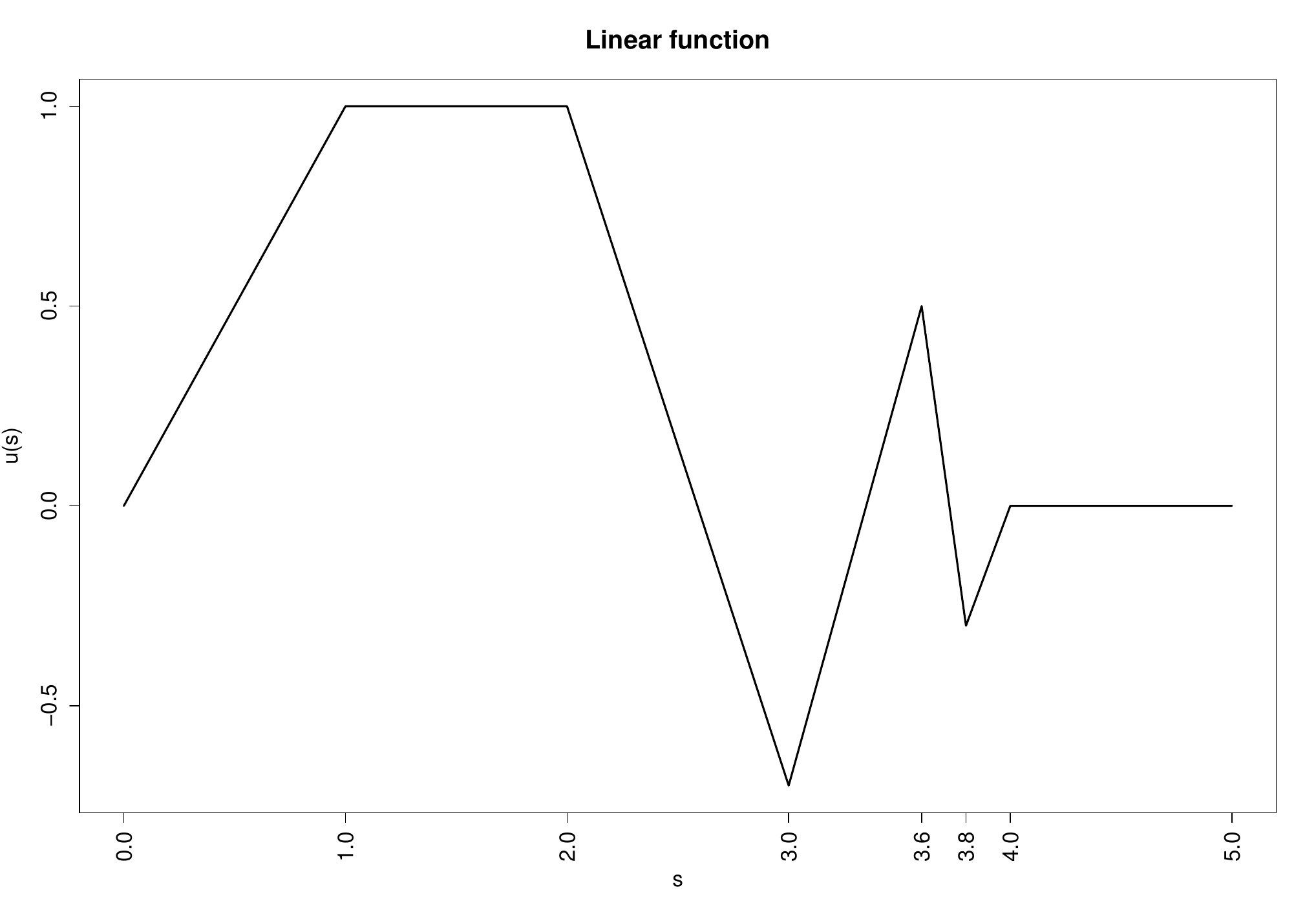}
	\includegraphics[width=.45\textwidth]{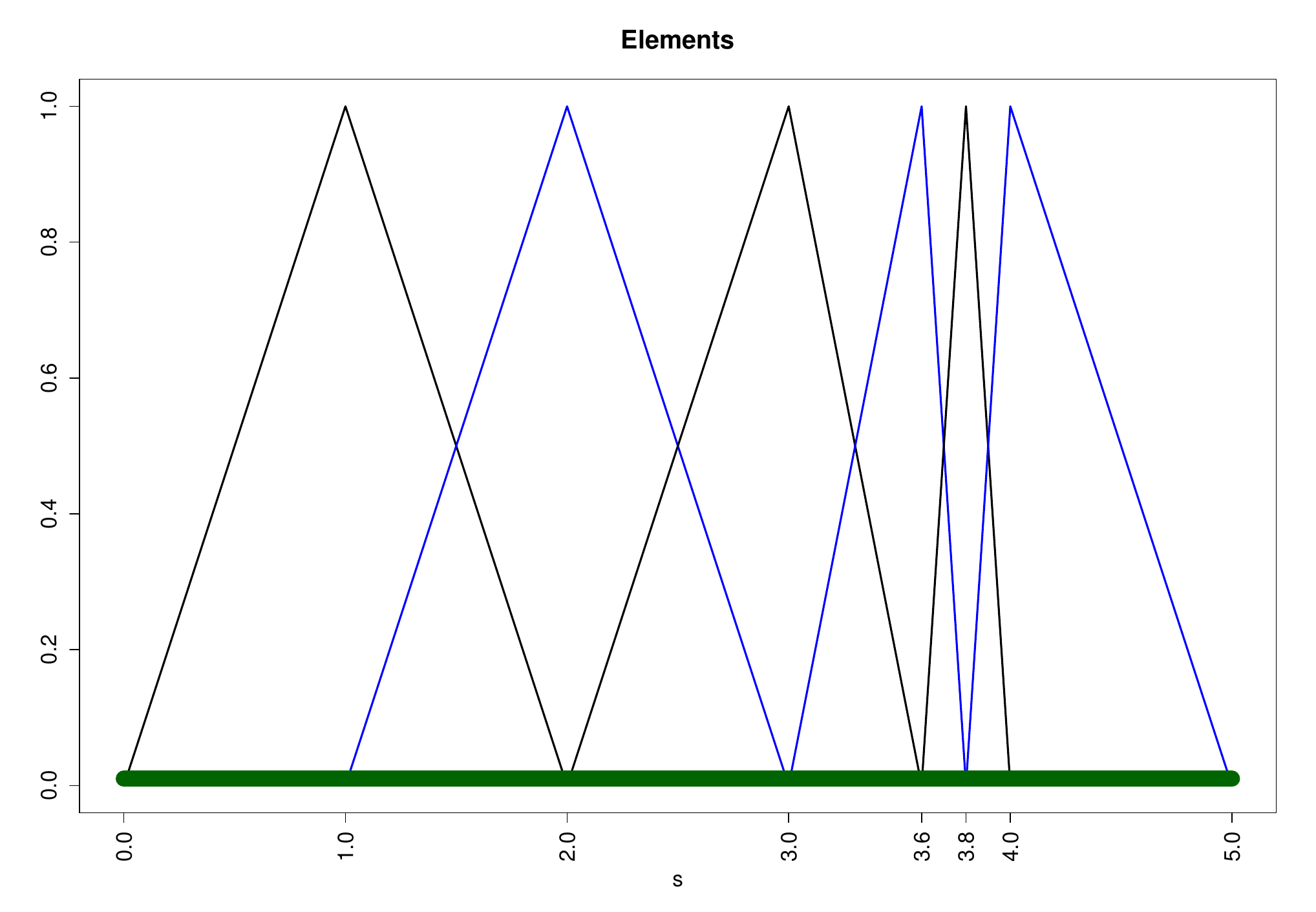}
	\includegraphics[width=.45\textwidth]{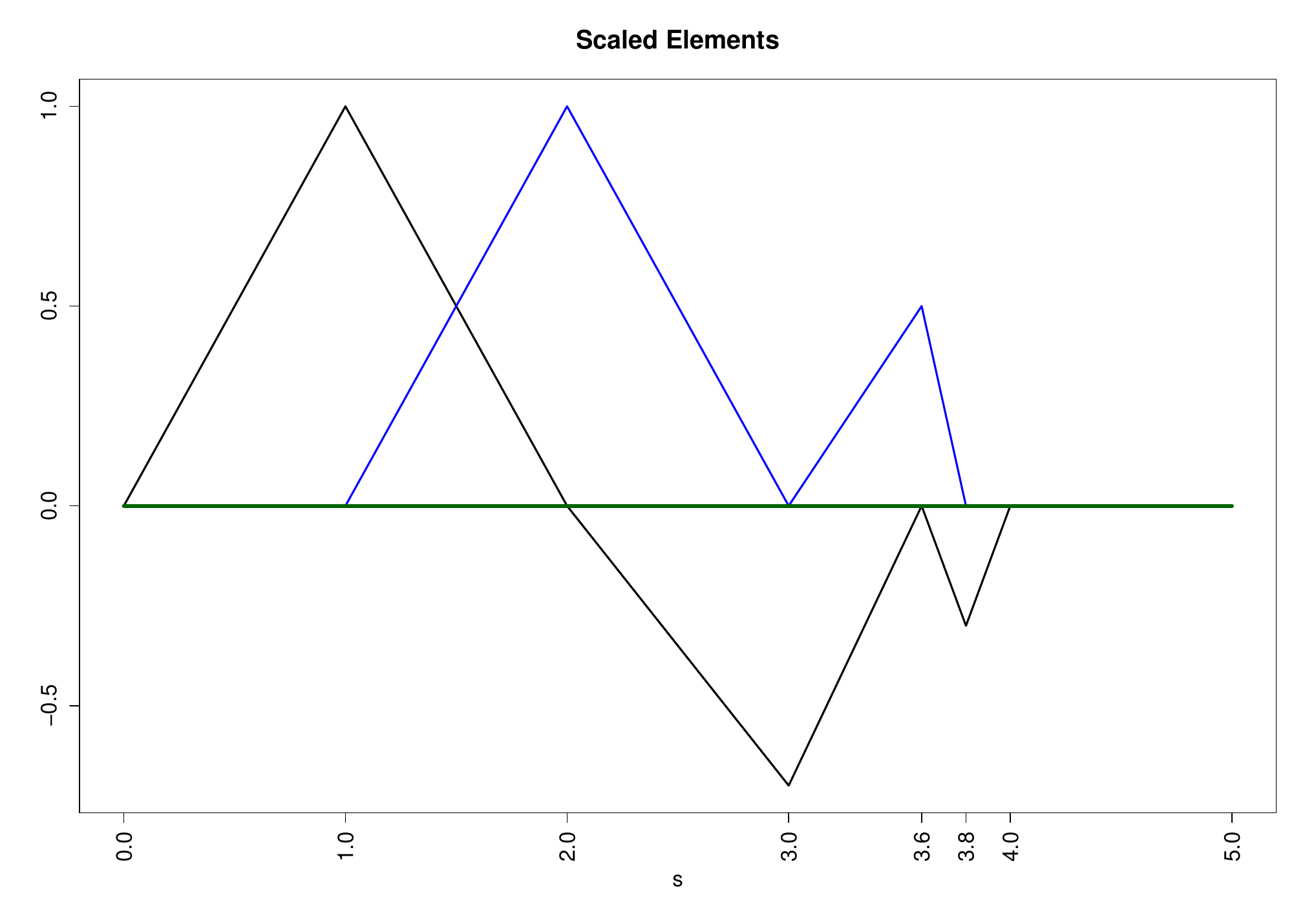}
	\includegraphics[width=.45\textwidth]{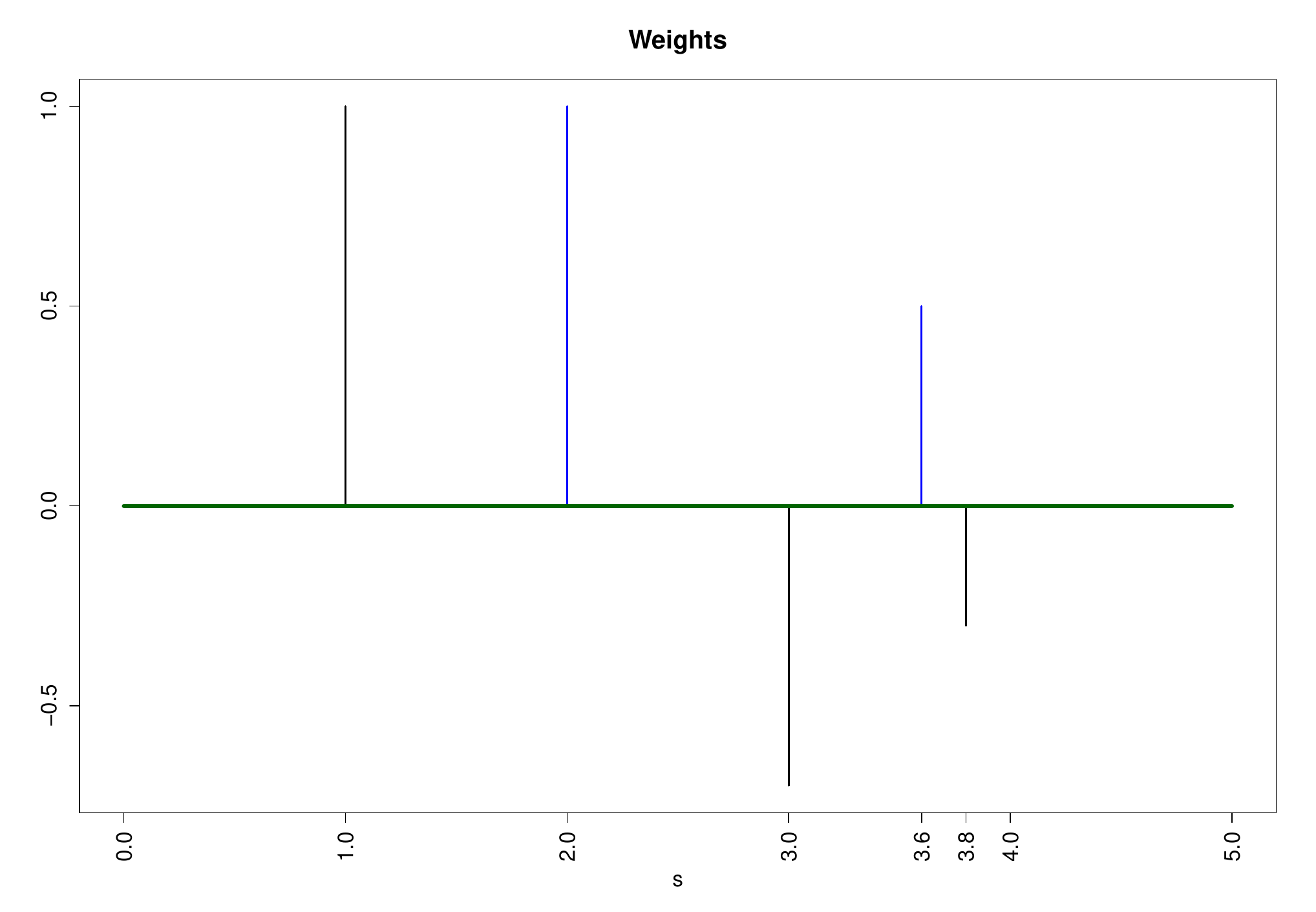}
	\caption{Linear approximations in 1D. 
		The first plot shows the piecewise linear function.
		The second shows the hat-functions (elements).
		Every hat-function is a separate function.
		Every second hat-function is coloured blue.
		The third plot is the decomposition of the linear function into a sum of scaled hat-functions. 
		Every second scaled hat function is marked in blue.
		Adding all these functions together gives the first plot.
		The fourth plot shows the weights/scales that connect plot 2 and plot 3.
		Starting with plot 2, you scale it with the weights in plot 4 to get plot 3.
		The chosen mesh nodes are the set of ticks along the x-axis.
		In total, plot 1 and plot 4 are equivalent representations of the same function.
	}
	\label{fig-linapprox1d}
\end{figure}

%\begin{frame}{Finite Element Method in 1D}{Principle: Any linear function is a sum of local hat-functions}
In this appendix we show how any linear function in 1 dimension can be written as a  sum of weighted elements, also known as a linear combination of elements.
See figure \ref{fig-linapprox1d} and the caption.

\end{document}